\documentclass[a4paper,twocolumn,floatfix]{revtex4-2}

\usepackage{graphicx}
\usepackage[utf8]{inputenc}
\usepackage[T1]{fontenc}
\usepackage{float}
\usepackage{amsmath,amssymb,amsfonts,amstext}
\usepackage{dsfont}

\usepackage[svgnames]{xcolor}
\usepackage{multirow}
\usepackage{enumitem}

\usepackage{threeparttable}
\usepackage{dcolumn}
\usepackage{booktabs}
\usepackage{caption}
\usepackage{float}
\usepackage{amsmath}
\usepackage{subcaption}

\begin{document}
\title{Probing Dark Energy Evolution Post-DESI 2024
}

\author{Lili Orchard $^1$,$^2$}
    \email{lili.orchard@duke.edu}
\author{V\'ictor H. C\'ardenas $^1$}
    \email{victor.cardenas@uv.cl}
  
    \affiliation{$^1$Universidad de Valparaíso, Instituto de Física y Astronomía, Valparaíso, Chile}
    \affiliation{$^2$ Department of Physics, Duke University, Durham, NC 27708, USA}

\begin{abstract}

We study the evidence for dark energy (DE) evolution at low redshift, using baryonic acoustic oscillations (BAOs) from the DESI Early Data Release, Pantheon+ Type Ia supernovae (SNe-Ia), and redshift space distortions (RSDs) to constrain cosmological parameters. Furthermore, we make use of the angular acoustic scale to analyse the effect of introducing condensed CMB information on the cosmological parameters informing DE evolution. The analysis is divided into cases based on the variability of priors inferred from early-time physics. Using a quadratic parametrisation, $X(z)$, for DE density, we find evidence for DE evolution in all cases, both with and without the angular acoustic scale. We reconstruct $X(z)$ using best fit parameters and find that DE density starts to exhibit dynamical behaviour at $z \sim 0.5 $, assuming negative values beyond $z\sim 1.5$. The data show no significant preference for $X(z)$CDM over a $\Lambda$CDM, with both models performing equally well according to our chosen metrics of reduced $\chi^2$ and the Durbin-Watson statistic. 

\end{abstract}

\date{\today} 

\maketitle

\section{Introduction}

The Early Data Release from the Dark Energy Spectroscopic Instrument (DESI, \cite{DESI:2024mwx, desicollaboration2024desi, desicollaboration2024desi2}) has motivated further exploration of dynamical dark energy (DE) in place of a cosmological constant in the standard cosmological model, $\Lambda$CDM \cite{gialamas2024interpreting, tada2024quintessential,Yuhang_2024,shlivko2024assessing,ramadan2024desi,lynch2024desi,notari2024consistent,wang2024dark, Chen:2024vuf, Colgain:2024ksa}. Prominant $\Lambda$CDM alternatives include quintessence models \cite{WETTERICH1988668, RatraPeebles, Frieman_1995, Turner_1997, Caldwell_1998, Steinhardt_1999} and theories of modified gravity \cite{Shankaranarayanan_2022, Tsujikawa_2010, Capozziello_2011, Boubekeur_2014, Johnson_2016, Clifton_2012, Nojiri_2017}, many of which have been revised in light of the recent DESI data release \cite{gialamas2024interpreting}. The most common approach to look for potential deviations from the $\Lambda$CDM model is to represent the DE component using the equation of state parameter $w(a)$. This approach assumes there is an unknown component with energy density given by
\begin{equation}
    \rho_{\text{DE}}(z) = \rho_{\text{DE}}{(0)} \exp \left[ 3\int_1^{1+z} (1+\omega(x)) d\ln x  \right].
\end{equation}
This approach, in principle, admits a phantom behavior ($\omega<-1$), but does not allow the excursion to negative DE densities. However, the authors of \cite{Maor:2000jy,Maor:2001ku}, demonstrate that the method is inherently constrained because the functions that connect observations with theory, such as the luminosity distance, rely on $\omega(z)$ through a complex integral relationship. This obscures detailed information about $\omega(z)$ and its temporal changes, reducing the precision with which $\omega(z)$ can be determined from existing data.

Rather than initially adopting a particular physical model for DE, given our lack of clear insight into its origin, in this work we opt for an exploratory approach, utilizing a probing function to represent the DE density. Our aim is to identify deviations from the $\Lambda$CDM model which assumes a constant DE component. This approach started in \cite{Wang:2001ht,Wang:2001da,Wang:2003cs}, where the authors used both a linear and a quadratic probe function for $X(z)=\rho_{DE}(z)/\rho_{DE}(0)$. For $\Lambda$CDM, $X(z)=1$, whilst if $X(z) \neq 1$ for {\it any} redshift $z$, this is an indication of DE evolution. In \cite{Cardenas:2014jya} it was found that low redshift observational data prefer that $X(z)$ decreases with $z$, even taking negative values for $z>1$. The same trend has been found using new data, such as the Pantheon sample \cite{Grandon:2021nls,Bernardo:2021cxi}, and also using new methods, such as non-parametric and even biology inspired ones, discussed in \cite{Bernardo:2021cxi,Bernardo:2022pyz}, with \cite{Bernardo:2022pyz} making use of the Pantheon+ sample. This trend of including the possibility of a negative $X(z)$ is a challenge for DE modeling. Furthermore, the authors of \cite{Malekjani:2023ple, Colgain:2024ksa} verify that this trend holds true when datasets (Pantheon+ and DES 5YR, respectively) are confined to redshift bins.

The paper is organised as follows. The next section introduces the methodology and datasets used for the statistical analysis. The main results are described in section III. We make our concluding remarks in section IV.

\section{Methodology and data}

Assuming an isotropic, homogeneous universe, characterised by a flat FLRW spacetime metric, the Hubble parameter is defined as follows, 
\begin{equation}\label{H(z)}
H(z)^2 = H_0^2\left[\Omega_\text{M}(1+z)^3 + \Omega_\text{R}(1+z)^4  + \Omega_\Lambda X(z)\right],
\end{equation}
where the energy densities parameters of matter, radiation, and dark energy are described by $\Omega_\text{M}$, $\Omega_\text{R}$, and $\Omega_\Lambda$, respectively. 
We restrict our analysis to the flat case, thus, $\Omega_\Lambda = 1 - (\Omega_\text{M} + \Omega_\text{R})$. Whilst $\Omega_\text{M}$ is a constant to be constrained directly, we make use of the relation,
\begin{equation}\label{H(z)}
\Omega_\text{R} = \Omega_\gamma + \Omega_\nu = (1 + 0.2271N_{\text{eff}})\Omega_\gamma,
\end{equation}
for $\Omega_\text{R}$ in our analysis of BAO data \cite{Lesgourgues_2012}. $N_{\text{eff}}$ is effective number of neutrinos, $\Omega_\gamma$ is photon energy density, and $\Omega_\nu$ is neutrino energy density. 

We adopt an agnostic approach in attempting to describe the behaviour of DE. Our analysis is not predicated on any physical assumptions about the nature of DE; instead, we choose to focus on dark energy as a purely geometric effect. As such, we devise a probe function with the capacity to detect dynamical dark energy if the data so prefer. Our probe function of choice is a second degree polynomial,
\begin{multline}\label{X(z)}
X(z) = \frac{x_0(z-z_1)(z-z_2)}{(z_0-z_1)(z_0-z_2)} + \frac{x_1(z-z_0)(z-z_2)}{(z_1-z_0)(z_1-z_2)} + \\
\frac{x_2(z-z_0)(z-z_1)}{(z_2-z_0)(z_2-z_1)},
\end{multline}
which is a natural non-linear deviation from unity, and advantageous over other parametrisations due to having few degrees of freedom. This parametrisation is a continuation of the work in \cite{Grandon:2021nls, Cardenas:2014jya, Bernardo:2021cxi}. Variables $x_0, x_1,$ and $ x_2$ are constant values of $X(z)$ evaluated at $z_0, z_1$, and $ z_2$, where $z_0<z_1<z_2$. Using the same definitions as in \cite{Grandon:2021nls}, $z_0 = 0$, $z_1 = z_\text{m}/2$, $z_2 = z_\text{m}$, where $z_\text{m} = 2.330$ is the maximum redshift value across the three datasets used in this analysis. The free parameters in $X(z)$ then become $x_1 = X(z_\text{m}/2)$ and $x_2 = X(z_\text{m})$. These definitions reduce (\ref{X(z)}) to 
\begin{equation}\label{X'(z)}
X(z) = 1 + \frac{z(4x_1 - x_2 -3)}{z_\text{m}} - \frac{2z^2(2x_1-x_2-1)}{z_\text{m}^2}.
\end{equation}
The first thing to note is the concordance between this cosmological model, which we shall refer to as $X(z)$CDM, and $\Lambda$CDM, in the case of no DE evolution. If the data prefer constant over dynamical DE density, then $x_1$ and $x_2$ will tend to unity, reducing (\ref{X'(z)}) to the $\Lambda$CDM model. $X(z)$CDM, therefore, may be thought of as an extension of the mostly successful $\Lambda$CDM model, in pursuit of exploring the possibility of dynamical DE at low redshift. 

The model is put under stress using four types of observational data: baryonic acoustic oscillation data from DESI \cite{DESI:2024mwx}, Type Ia supernovae from the Pantheon+ sample \cite{Scolnic_2022}, redshift space distortions from a sample compiled by the authors of \cite{Kazantzidis_2018}, and cosmic microwave background radiation data from the acoustic angular scale, also used in \cite{DESI:2024mwx}.

\subsection{Baryonic Acoustic Oscillations}

The first dataset we use to extract cosmological parameters is DESI's first year release of baryonic acoustic oscillations (BAOs) \cite{DESI:2024mwx}. The dataset consists of 12 points, with redshift ($z_{\text{eff}}$) ranging from 0.295 to 2.330. There are five redshift values at which both the transverse ($D_\text{M}$/$r_\text{d}$) and radial ($D_\text{H}$/$r_\text{d}$) components of BAOs are reported, and two redshifts at which only the an angle-averaged distance, $D_\text{V}$/$r_\text{d}$, is reported.

The theoretical values for these BAO measurements are obtained using the following relations, where $r_\text{d}$ the drag-epoch sound horizon:
\begin{equation}\label{D_H}
\frac{D_\text{H}(z)}{r_\text{d}} = \frac{c}{H(z)r_\text{d}}
\end{equation}
in the radial direction,
\begin{equation}\label{D_M}
\frac{D_\text{M}(z)}{r_\text{d}} = \frac{c}{r_\text{d}}\int_{0}^{z}\frac{dz'}{H(z')}
\end{equation}
in the transverse direction, and 
\begin{equation}\label{D_v}
\frac{D_\text{V}(z)}{r_\text{d}}= \frac{(zD_\text{M}(z)^2D_\text{H}(z))^\frac{1}{3}}{r_\text{d}}
\end{equation}
for the angle-averaged distance, calculated instead for the two data points with a significantly low signal-to-noise ratio \cite{desicollaboration2024desi}.

To constrain the parameters of a given cosmological model using BAO data, we require prior knowledge of a sort to break the degeneracy between $r_\text{d}$ and $H_0$. In \cite{DESI:2024mwx} this was achieved in two ways, the first of which makes use of CMB \cite{Planck:2018} and CMB lensing \cite{Carron_2022, Madhavacheril_2024, maccrann2023atacama, Qu_2024} information to directly calibrate $r_\text{d}$, such that $r_\text{d} = 147.09 \pm 0.26 $ Mpc is fixed. The second method involves introducing a Big Bang Nucleosynthesis (BBN)-inferred prior on $\Omega_\text{b} h^2$, followed by use of the early-time physics approximation for the drag-epoch sound horizon \cite{Brieden_2023}. 

We opt for a different approach to the $r_\text{d}$-$H_0$ degeneracy problem that allows for an explicit dependence on the parameters of a given cosmological model in the determination of $r_\text{d}$. This approach is similar to the second method outlined above, however, instead of using an approximate equation for $r_\text{d}$, we used the integral definition to calculate the drag-epoch sound horizon, 
\begin{equation}\label{rd}
r_\text{d} = \int_{z_\text{d}}^{\infty }\frac{c_\text{s}(z)}{H(z)} dz,
\end{equation}
where $z_\text{d}$ is redshift at the drag-epoch, and $c_\text{s}(z)$, the speed of sound, is given by
\begin{equation}\label{c_s}
c_\text{s}(z) = \frac{c}{\sqrt{\left ( 1 + \frac{3\rho_\text{B}(z)}{4\rho_\gamma(z)}\right )}}.
\end{equation}
In this way, $r_\text{d}$ is treated as another theoretical function, defined in terms of cosmological parameters, to be fitted to data. 

It should be noted that the adoption of $r_\text{d}$ in integral form alone is not sufficient to overcome $r_\text{d}$-$H_0$ degeneracy and determine the remaining cosmological parameters. The integral approach is also dependent on prior knowledge, specifically, the calculation of $r_\text{d}$ involves $N_{\text{eff}}$, $\omega_{\text{b}} = \Omega_\text{b} h^2$, where $h = H_0/100$, and $z_\text{d}$. We examine the impact on our results of assuming such values to be fixed, versus allowing them to be free variables within the equation for $r_\text{d}$. When fixed, the variables assume values of $N_{\text{eff}} = 3.044$ (standard model prediction \cite{Bennett_2021, Mangano:2005cc}), $\omega_\text{b} = 0.02218$ \cite{schöneberg20242024}, and $z_\text{d} = 1059$ \cite{Planck:2015fie}. 

In the case of $N_{\text{eff}}$, $\omega_\text{b}$, and $z_\text{d}$ as free variables, we introduce Gaussian priors on all three variables. The prior on $\omega_\text{b}$ is BBN-inferred \cite{schöneberg20242024}, and the prior on $z_\text{d}$ is a Gaussian constructed around a mean of value 1059, with standard deviation 10-- a conservative estimate. The authors of \cite{DESI:2024mwx} also consider varying $N_{\text{eff}}$ in their analysis. Combining CMB and CMB lensing information from \textit{Planck} \cite{Planck:2018} and the Atacama Cosmology Telescope (ACT) \cite{Madhavacheril_2024, Qu_2024, maccrann2023atacama} with DESI BAO data, they report a value of $N_{\text{eff}} = 3.10 \pm 0.17$. We use this result for our $N_{\text{eff}}$ prior.

Allowing these parameters to vary serves primarily as an exercise in agnosticism. The fixed values for all three variables are derived using $\Lambda$CDM, introducing a dependence on the very model we aim to explore deviations from (albeit deviations at low redshift). Whilst we expect that an alternative cosmological model will agree with the results of $\Lambda$CDM at early times, it is nonetheless interesting to explore the implications of allowing $N_{\text{eff}}$, $\omega_\text{b}$, and $z_\text{d}$ to become free variables. 

Returning to our analysis of BAOs, in equations (\ref{D_H}), (\ref{D_M}), and (\ref{D_v}), $r_\text{d}$ is no longer a constant but rather a function of cosmological parameters. A prior is never imposed on $r_\text{d}$ explicitly; we opt only to impose priors on variables involved in its calculation.

\subsection{Type Ia Supernovae}

The second and largest dataset used in our analysis is the Pantheon+ Type Ia SNe (SNe-Ia) sample, consisting of 1701 observed magnitude measurements at redshifts ranging from 0.00122 to 2.26137 \cite{Scolnic_2022}. Of these data, 77 data points are calibrated using Cepheid variables, and the remaining 1624 points are left uncalibrated. Making use of the luminosity distance equation,
\begin{equation}\label{D_L}
D_\text{L}(z) = c \cdot (1+z) \int_{0}^{z}\frac{dz'}{H(z')},
\end{equation}
with explicit dependence on the Hubble parameter, allows for theoretical apparent magnitude to be expressed as follows,
\begin{equation}\label{m}
m = 5 \text{log}\left (D_\text{L}(z)/h\right ) + M,
\end{equation}
where $M$ is the absolute magnitude of SNe-Ia.
Best fit values for the cosmological parameters in $H(z)$ may be fitted to SNe-Ia data by comparing apparent magnitude predictions from (\ref{m}) with the Pantheon+ results. Since the SNe-Ia sample is used in combination with BAO data throughout this analysis, we omit the use of a standard candle-inferred prior on $H_0$, such as the Riess prior (R21) of $H_0^{R21} = 73.04 \pm 1.04$ km s$^{-1}$ Mpc$^{-1}$ \cite{Riess_2022}. At no point in this analysis is $H_0$ subject to priors of any kind, in tandem with the motivation for seeking a model beyond $\Lambda$CDM: uncertainty around the true value of $H_0$.

\subsection{Redshift Space Distortions}
The third data included in this analysis are 63 growth rate parameter values ($f\sigma_8$) from redshift space distortions (RSDs) at redshifts ranging from 0.001 to 1.944. 
In this work we use a data set larger than other studies. The problem with this sample, discussed in \cite{Adil:2023jtu}, is that these data points are correlated due to overlaps in the galaxy samples used for their derivation, which ideally necessitates the use of a large covariance matrix for combined analysis. However, a complete covariance matrix for this dataset or its subsets is not available in the literature. This lack of a full covariance matrix poses a challenge, as using arbitrary subsamples \cite{Nguyen:2023fip} of the full $f\sigma8$ dataset can result in the loss of valuable information \cite{Kazantzidis_2018}. For this reason we decided to use the full set.

This dataset is an amalgamation of RSD data collected between 2006 and 2008, and was first constructed by the authors of \cite{Kazantzidis_2018} in their analysis of the $f\sigma_8$ tension. This tension can also be related to the $S_8$ tension, from weak lensing measurements, viewed in the plane $S_8-\Omega_M$, where $S_8=\sigma_8 \sqrt{\Omega_M/0.3}$. Weak lensing measurements from three galaxy surveys, the Dark Energy Survey (DES), Kilo-Degree Survey (KiDS), and Hyper Suprime-Cam Subaru Strategic Program (HSC SSP), have yielded consistent results, revealing a $2\text{ - }3\sigma$ tension with \textit{Planck} \cite{DES:2021wwk,KiDS:2020suj,Dalal:2023olq}.

The following second-order ordinary differential equation describes the evolution of density perturbations,
\begin{equation}\label{ODE}
\delta(a)''+\left(\frac{3}{a}+ \frac{E'(a)}{E(a)} \right)\delta(a)' - \frac{3}{2} \frac{\Omega_{0m}}{a^5E(a)^2} \delta(a) = 0,
\end{equation}
where $E(a)$ is the normalized Hubble parameter as a function of the scale factor ($a = \frac{1}{1+z}$). In a similar manner to the authors of \cite{Kazantzidis_2018}, we solve this ODE numerically for $ 0 < a < 1$ to generate a function for $f\sigma_8$ \cite{Percival_2005} in terms of cosmological parameters used in the analysis of BAO and SNe-Ia data. 

This sample, however, lacks the accuracy of the previous two datasets due to significantly larger errors associated with the $f\sigma_8$ measurements. Thus, whilst alone it has a limited ability to precisely determine cosmological parameters, in conjunction with the other two datasets, the RSD data serve both to constrain the model further, and as a consistency check for our results that do not involve RSDs.

As the authors of \cite{Kazantzidis_2018} highlight, smaller RSD samples are typically seen in cosmological analyses, works \cite{PhysRevD.96.023542, Nesseris:2007pa, Nesseris_2015,Basilakos:2014yda, Basilakos_2016, Basilakos_2017, P_rez_Romero_2018, 2016MNRAS.456.3743A, G_mez_Valent_2017, G_mez_Valent_2018, Mehrabi_2015,Baker_2014,Pouri:2014nta,Paul_2013,LHuillier:2017ani} being examples from prior to the publication of \cite{Kazantzidis_2018}. More recently, this still holds true in papers exploring $\Lambda$CDM alternatives using combinations of BAO, SNe-Ia, and RSD data \cite{Cardona:2022pwm, Cardona_2024, Arjona:2020yum,Calderon:2022cfj, Garcia_Quintero_2019}, with a reluctance to use all available RSD datasets in aggregate due to correlated data points. Since the total available RSD data lack a full covariance matrix to account for these correlations, the authors of \cite{Kazantzidis_2018}  introduce "a nontrivial covariance matrix correlating
randomly 20 $\%$ of the RSD data points." They report that this has no significant effect on the main results of their analysis, hence, we proceed with the full dataset (see Table II of \cite{Kazantzidis_2018}).

It is worth noting a difference in parametrisation of $H(z)$ in the analysis of BAO and RSD data. In the calculation of $r_\text{d}$ from (\ref{rd}), it is necessary to account for the radiation contribution to (\ref{H(z)}) as a distinct entity, in alignment with early-time physics. Alternatively, in our analysis of the late-time RSD data, we treat $\Omega_\text{R}$ as negligible.

\subsection{Acoustic Angular Scale}

The final constraint imposed on the parameters of each cosmological model is the CMB-inferred acoustic angular scale from \textit{Planck} \cite{Planck:2018}. We incorporate this condensed CMB information into our analysis in the same manner as the authors of \cite{DESI:2024mwx}. The acoustic angular scale, $\theta_*$, is given by $\theta_* = r_*/ D_\text{M}(z_*)$, where 
\begin{equation}\label{theta}
r_* = \int_{z_*}^{\infty}\frac{c_\text{s}(z)}{H(z)} dz,
\end{equation}
and $z_* \sim 1090$ is the redshift at the epoch of recombination \cite{Planck:2018}. $\theta_*$ may be interpreted geometrically as "the BAO scale imprinted in the CMB anisotropies at recombination," \cite{DESI:2024mwx}. As such, introducing $\theta_*$ condenses CMB-realised information about the early universe into a constraint on the late-time ($z \leq 2.33$) data from DESI, Pantheon+, and RSDs. Whilst the \textit{Planck} value of $100\theta_* = 1.04109 \pm 0.00030$ makes use of the $\Lambda$CDM model in its derivation, here we invoke our assumption that $\Lambda$CDM alternatives ought to agree with $\Lambda$CDM at early times. We leave $\theta_*$ as a function of cosmological model parameters with a \textit{Planck}-inferred Gaussian prior of mean and standard deviation as cited above.

\begin{table}[htbp]
    \centering
    \begin{tabular}{@{}ll@{}}
        \toprule
        \textbf{Parameter} & \textbf{Prior} \\
        \midrule
        $\omega_\text{b}$ & $\mathcal{N}(\mu = 0.02218, \sigma = 0.00055)$ \\
        $N_{\text{eff}}$ & $\mathcal{N}(\mu = 3.10, \sigma = 0.17)$ \\
        $z_\text{d}$ & $\mathcal{N}(\mu = 1059, \sigma = 10)$ \\
        $\theta_*$ & $\mathcal{N}(\mu = 1.04109, \sigma = 0.00030)$\\
        \bottomrule
    \end{tabular}
    \caption{Gaussian priors used in analysis.}
    \label{tab:prior_values}
\end{table}

\subsection{Calculating $\chi_{\text{red.}}^2$ and Durbin-Watson}

For both $\Lambda$CDM and $X(z)$CDM, we report best fit cosmological parameters for the combinations of DESI, Pantheon+, and RSD data, with and without $\theta_*$ (see Table \ref{tab:1}). Below is outlined our method for the calculation of reduced chi-squared ($\chi_{\text{red.}}^2$) and the Durbin-Watson (DW) statistic, measures of goodness of fit. 

Considering the DESI sample alone, the 10 data points for which either $D_\text{M}/r_\text{d}$ or $D_\text{H}/r_\text{d}$ is reported have associated covariance matrices ($C_\text{D}$). We use 

\begin{equation}\label{chiDESI1}
\begin{aligned}
\chi_{\text{D},1}^2 &= \sum_{i=1}^{5} (\mathbf{D}(z_i)^{th} - \mathbf{D}(z_i))^T C_{\text{D}, i}^{-1} (\mathbf{D}(z_i)^{th} - \mathbf{D}(z_i)),
\end{aligned}
\end{equation}
where $\mathbf{D}(z_i)$ is the vector ($D_\text{M}(z_i)/r_\text{d}$, $D_\text{H}(z_i)/r_\text{d}$) and $\mathbf{D}(z_i)^{th}$ is its theoretical counterpart, to calculate the contribution of these data to the final DESI chi-squared statistic, $\chi_{DESI}^2$. For the remaining two data points, $D_\text{V}/r_\text{d}$ is reported with an uncertainty value, $\sigma$. We use 

\begin{equation}\label{chiDESI2}
\chi_{\text{D},2}^2 = \sum_{i=1}^{2} \frac{(D_\text{V}(z_i)-D_\text{V}(z_i)^{th})^2}{\sigma_i^2},
\end{equation}
to estimate their contribution to $\chi_{\text{DESI}}^2$. Thus, for the entire BAO sample, we have $\chi_{\text{DESI}}^2 = \chi_{\text{D},1}^2 + \chi_{\text{D},2}^2$.

For Pantheon+, $\chi_{\text{SN}}^2$ is computed using

\begin{equation}\label{chiSN1}
\chi_{\text{SN}}^2 = (\mathbf{\Delta\mu})^T C_{\text{SN}}^{-1} (\mathbf{\Delta\mu}),
\end{equation}
where $C_{\text{SN}}$ is the associated covariance matrix. The components of vector $\mathbf{\Delta\mu}$ are given by $ \mu(z_i)- \mu(z_i)^{th}$ for the 1624 uncalibrated data points, and $\mu(z_i) - M - \mu_{\text{Ceph}}(z_i)$ for the remaining 77 apparent magnitude values calibrated using Cepheid variables.

For the RSD data,
\begin{equation}\label{chiRDS}
\chi_{\text{RSD}}^2 = \sum_{i=1}^{63}\frac{(f\sigma_8(z_i)-f\sigma_8(z_i)^{th})^2}{\sigma_i^2}.
\end{equation}
We weight each sample's contribution to the final chi-squared statistic evenly, such that

\begin{equation}\label{chitot}
\chi^2 = \chi_{\text{DESI}}^2 + \chi_{\text{SN}}^2 + \chi_{\text{RSD}}^2.
\end{equation}
With the inclusion of $\theta_*$, 

\begin{equation}\label{chitot}
\chi^2 = \chi_{\text{DESI}}^2 + \chi_{\text{SN}}^2 + \chi_{\text{RSD}}^2 + \chi_{\theta_*}^2,
\end{equation}
where $\chi_{\theta_*}^2$ assumes the mean and standard deviation of $\mathcal{N}(\theta_*$) as the measured value (see Table \ref{tab:prior_values}). In the cases where $N_\text{eff}, \omega_\text{b}$, and $z_\text{d}$ are variable, their associated Gaussian priors (see Table \ref{tab:prior_values}) are also added explicitly to $\chi^2$.

To obtain the reduced $\chi^2$ statistic, we use
\begin{equation}\label{chired}
\chi_{\text{red.}}^2 = \frac{\chi^2}{K},
\end{equation}
where the number of degrees of freedom, $K$, is assumed equal to $N - P$ for $N$ data points and $P$ parameters \cite{andrae2010dos}. We treat $\theta_*$ as a single data point in its contribution to $N$.

The DW statistic is given by
\begin{equation}\label{dw}
\text{DW} = \frac{\sum_{i=2}^{N} (e_i - e_{i-1})}{\sum_{i=1}^{N} e_i^2},
\end{equation}
where $e_i$ is the residual for a given data point; $e_i=d_i-m_i$ where $d_i$ is the data point and $m_i$ is the corresponding model value. For a large number data points, the statistic is equal to $2(1-\hat{\rho})$, where $\hat{\rho}$ is the autocorrelation of the residuals. 

The Durbin-Watson (DW) statistic assesses autocorrelation in the residuals of a regression model. It ranges from 0 to 4, with a value near 2 indicating no autocorrelation, values less than 2 suggesting positive autocorrelation, and values greater than 2 indicating negative autocorrelation. Positive autocorrelation (DW$ < 2$) implies underestimated data variability, leading to potentially biased inferences, while negative autocorrelation (DW$ > 2$) suggests overestimated variability. Evaluating the DW statistic helps verify whether residuals are independent and identically distributed, a key model assumption, and informs adjustments such as adding variables or altering estimation methods. We use the Python module \texttt{statsmodels} \cite{seabold2010statsmodels} to compute the statistic.

\subsection{Markov Chain Monte Carlo (MCMC) Sampling }

We make use of the public code \texttt{emcee} \cite{Foreman-Mackey:2013} for MCMC sampling. As outlined in \cite{Bernardo:2022pyz}, the Metropolis-Hastings algorithm is invoked to attain the posterior for the relevant cosmological parameters from the likelihood function and prior distribution of parameters. We minimize $\chi^2$ using the Nelder-Mead method to decipher a starting position within the parameter space, before walkers are set off for 5000 steps. The first 20 samples are neglected as burn in. We use \texttt{GetDist} \cite{lewis2019getdist} to visualise the triangle plots for the resulting posterior distributions.

\section{Results}

\newcommand{\ra}[1]{\renewcommand{\arraystretch}{#1}}

\begin{table*}
    \centering
    \ra{1.3}
    \setlength{\tabcolsep}{4pt} 
    \begin{subtable}{\linewidth}
    \begin{tabular}{@{}p{3.5cm}rrrrrr||rr@{}}
        \toprule
        & $\Omega_{\text{M}}$ & $H_0$ & $M$ & $\sigma_8$ & $x_1$ & $x_2$& $\chi^2_\text{red.}$ & DW \\ 
        \midrule
        \textbf{Flat $\Lambda$CDM} & & & & & & \\
        
        DESI+Panth.+RSD & $0.311 \substack{+0.017 \\ -0.015}$ & $70.59 \substack{+0.75 \\ -0.79}$ &  $-19.34 \substack{+0.02 \\ -0.03}$ & $0.757 \substack{+0.020 \\ -0.022}$ & $\text{--}$ & $\text{--}$ & 0.895 &  1.68 \\

        DESI+Panth.+RSD+$\theta_*$ &  $0.286 \substack{+0.006\\ -0.005}$ & $69.35 \substack{+0.40\\ -0.43}$ &   $-19.39 \pm 0.01$ & $0.771 \substack{+0.015\\ -0.013}$ & $\text{--}$ & $\text{--}$  & 0.902  &  1.70 \\

        \textbf{Flat $X(z)$CDM} & & & & & & \\

        DESI+Panth.+RSD & $0.348 \pm 0.014$ & $73.38 \substack{+0.89 \\ -0.85}$ &  $-19.25 \substack{+0.03 \\ -0.02}$ & $0.721 \substack{+0.015 \\ -0.016} $ & $0.37 \substack{+0.13 \\ -0.15}$ & $-1.69 \substack{+0.48 \\ -0.74}$ &  0.886  & 1.72 \\

        DESI+Panth.+RSD+$\theta_*$ &  $0.335 \substack{+0.013 \\ -0.011}$ & $72.33 \substack{+0.63 \\ -0.80}$ &   $-19.28 \pm 0.02$ & $0.731 \substack{+0.015 \\ -0.017}$ & $0.42 \substack{+0.14 \\ -0.13}$  & $-1.74 \substack{+0.60 \\ -0.59}$ &   1.14  & 1.74\\

        \bottomrule
    \end{tabular}
    \caption{Fixed $N_{\text{eff}} = 3.044$, $\omega_{{\text{b}}} = 0.02218$, z$_{\text{d}} = 1059$}
    \label{tab:1}
    \end{subtable}
 
    \vspace{10pt}

    \begin{subtable}{\linewidth}
    \begin{tabular}{@{}p{3cm}rrrrrr||rr@{}}
        \toprule
        & $\Omega_{\text{M}}$ & $H_0$ & $M$ & $\sigma_8$ & $x_1$ & $x_2$ & $\chi^2_\text{red.}$ & DW \\ 
        \midrule
        \textbf{Flat $\Lambda$CDM} & & & & & & &\\
  
        DESI+Panth.+RSD & $0.313 \substack{+0.010 \\ -0.011}$ & $71.64 \substack{+0.69 \\ -0.72}$  &  $-19.31 \pm 0.02 $ &  $0.755 \pm 0.014$ & $\text{--}$ & $\text{--}$ & 0.893 & 1.69 \\

        DESI+Panth.+RSD+$\theta_*$ & $0.298 \substack{+0.005 \\ -0.006}$ & $71.36 \pm 0.72 $  &  $-19.32 \pm 0.02 $ &  $0.766 \pm 0.015$ & $\text{--}$ & $\text{--}$ & 1.19 & 1.71 \\

        \textbf{Flat $X(z)$CDM} & & & & & & &\\
        
        DESI+Panth.+RSD & $0.350 \pm 0.014$ & $73.24 \substack{+0.92 \\ -0.84}$  &  $-19.25 \substack{+0.03 \\ -0.02} $ &  $0.722  \substack{+0.016 \\ -0.015}$ & $ 0.40\substack{+0.15 \\ -0.17} $ & $-1.64\substack{+0.61 \\ -0.68}$ & 0.887 & 1.72\\

        DESI+Panth.+RSD+$\theta_*$ & $0.336 \substack{+0.010 \\ -0.011}$ & $72.67 \substack{+0.84 \\ -0.92}$  &  $-19.27 \substack{+0.02 \\ -0.03} $ &  $0.730  \substack{+0.016 \\ -0.014}$ & $ 0.47\substack{+0.16 \\ -0.14} $ & $-1.56\substack{+0.68 \\ -0.52}$ & 1.11 & 1.74\\
        \bottomrule
    \end{tabular}
    \caption{Variable $N_{\text{eff}} = 3.10 \pm 0.17$, $\omega_{\text{b}} = 0.02218 \pm 0.00055$; fixed z$_{\text{d}} = 1059$}
    \label{tab:2}
    \end{subtable}
 
    \vspace{10pt}

    \begin{subtable}{\linewidth}
    \begin{tabular}{@{}p{3.5cm}rrrrrr||rr@{}}
        \toprule
        & $\Omega_{\text{M}}$ & $H_0$ & $M$ & $\sigma_8$ & $x_1$ & $x_2$ & $\chi^2_\text{red.}$ & DW \\ 
        \midrule
        \textbf{Flat $\Lambda$CDM} & & & & & & & &\\

        DESI+Panth.+RSD  &  $0.310 \substack{+0.012 \\ -0.011}$ & $72.10 \substack{+0.73 \\ -0.65}$ &   $-19.29 \pm 0.02$ & $0.756 \pm 0.016$ & $\text{--}$ & $\text{--}$ & 0.893 & 1.70\\
        
        DESI+Panth.+RSD+$\theta_*$ &  $0.295 \substack{+0.008 \\ -0.007}$ & $71.44 \substack{+0.56\\ -0.51}$ &   $-19.32 \pm 0.02$ & $0.768 \pm 0.014$ & $\text{--}$ & $\text{--}$ & 1.12 & 1.71 \\
    
        \textbf{Flat $X(z)$CDM} & & & & & & & &\\

        DESI+Panth.+RSD  &  $0.352 \substack{+0.016 \\ -0.014}$ & $73.38 \substack{+0.92 \\ -0.89}$ &   $-19.25 \substack{+0.03 \\ -0.02}$ & $0.720 \substack{+0.015 \\ -0.016} $ & $0.35\substack{+0.20 \\ -0.14} $ & $-1.91 \substack{+0.89 \\ -0.34} $ & 0.889 & 1.72 \\
        
        DESI+Panth.+RSD+$\theta_*$ &  $0.343 \substack{+0.012 \\ -0.015}$ & $72.80 \substack{+0.71 \\ -0.96}$ &   $-19.26 \substack{+0.02 \\ -0.03}$ & $0.725 \substack{+0.018 \\ -0.016} $ & $0.37\substack{+0.23\\ -0.13} $ & $-2.1 \substack{+1.0 \\ -0.3} $ & 1.17 & 1.74\\

        \bottomrule
    \end{tabular}
    \caption{Variable $N_{\text{eff}} = 3.10 \pm 0.17$, $\omega_{\text{b}} = 0.02218 \pm 0.00055$, z$_{\text{d}} = 1059 \pm 10$ }
    \label{tab:3}
    \end{subtable}
 
    \vspace{10pt}

\caption{Results from MCMC sampling. We report results for two cosmological models: $X(z)$CDM and $\Lambda$CDM, in two cases: 1) DESI+Panth.(SNe-Ia)+RSD, and 2) DESI+Panth.(SNe-Ia)+RSD+$\theta_*$. Two measures of goodness of fit, $\chi^2_\text{red.}$ and DW, are shown in the final two columns. Prior knowledge used to break $r_\text{d}$-$H_0$ degeneracy given in subcaptions.}

\end{table*}

Tables \ref{tab:1}, \ref{tab:2}, and \ref{tab:3} summarise the results of MCMC sampling in the three different cases with varying constraints on $N_{\text{eff}}$, $\omega_{\text{b}}$, and $z_{\text{d}}$. We begin the analysis of our results by discussing the goodness of fit of both cosmological models. We proceed to analyse the effect of varying $N_{\text{eff}}$, $\omega_{\text{b}}$, and $z_{\text{d}}$ on the resulting cosmological parameters for both models. We compare $\Lambda$CDM and $X(z)$CDM values for $H_0$ and $\sigma_8$ with previous analyses. Finally, we reconstruct $X(z)$ and $q(z)$, the deceleration parameter, using best fit parameters, and explore the implications for DE evolution.

In our evaluation of goodness of fit we make use of two statistics: $\chi_\text{red.}^2$ and Durbin-Watson (DW). The combination allows for a more robust analysis, as the two statistics assess different qualities of a given model. The DW statistic tests for serial autocorrelations of the residuals, a characteristic that $\chi_\text{red.}^2$ cannot account for \cite{BeyondChiSquared}, with a desired value of 2. The more familiar statistic within the field of cosmology, $\chi_\text{red.}^2$, has an ideal value of 1, with lower values implying over-fitting \cite{andrae2010dos}. 

The statistics in Table \ref{tab:1}, when considered in aggregate, imply no significant preference for $X(z)$CDM over $\Lambda$CDM. For the combination of all three datasets with $\theta_*$, the $\chi_\text{red.}^2$ and DW values for $X(z)$CDM are 1.14 and 1.74, respectively. In the same case, for $\Lambda$CDM, $\chi_\text{red.}^2 = 0.902$ and DW $=1.70$. The proximity of the $\chi_\text{red.}^2$ and DW values to their desired values of 1 and 2, respectively, indicates successful data-fitting from both models. The case outlined above is the most constrained (fewest free parameters) and hence, shall be a focal point in our analysis. However, the success of $X(z)$CDM being comparable to that of $\Lambda$CDM extends to all cases. Furthermore, the DW statistic favours $X(z)$CDM universally. It should also be highlighted that all $\chi_\text{red.}^2$ and DW values across the three tables remain acceptably close to the desired values. In this way, both models are competitive in describing the observational data used, both fitting the data and obtaining a small autocorrelation. Hence, neither model should be outright rejected.

By a comparison of the results in Tables \ref{tab:1}, \ref{tab:2}, and \ref{tab:3} we observe -- as expected -- that increasing the number of free parameters increases the resulting best fit value for $H_0$ in the case of $\Lambda$CDM: DESI+Panth.+RSD+$\theta_*$. We find that $H_0$ for $N_{\text{eff}}$, $\omega_{\text{b}}$, and $z_{\text{d}}$ fixed lies $3.1\sigma$ lower than $H_0$ for $N_{\text{eff}}$, $\omega_{\text{b}}$, and $z_{\text{d}}$ all variable. We define $\sigma = \sqrt{\sigma_1^2 + \sigma_2^2}$, where $\sigma_1$ and $\sigma_2$ are the mean uncertainty values associated with the given results. This implies that increased uncertainty in early-time physics variables and, hence, fewer assumptions in our $\Lambda$CDM model, does impact best-fit cosmological parameters, specifically increasing $H_0$. We note that in both cases with $N_{\text{eff}}$ variable (Tables \ref{tab:3} and \ref{tab:2}), a consequence of this variability is $N_\text{eff}$ increasing to $3.4 \pm 0.1$, a $1.5\sigma$ deviation from the prior distribution mean. This increase in cosmological parameters is not observed in the corresponding $X(z)$CDM results to a significant degree.

\begin{figure}
\centering
\includegraphics[width=0.5\textwidth]{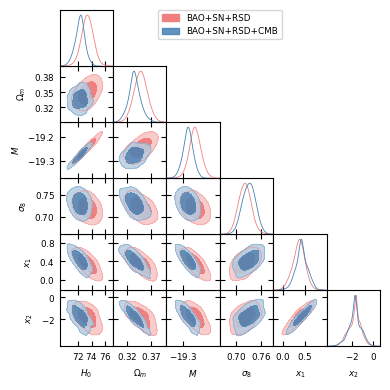}
\caption{\label{fig:triangleplot} Posteriors of the parameters in  $X(z)$CDM analysis with $N_{\text{eff}}$, $\omega_{\text{b}}$, and $z_{\text{d}}$ fixed (Table \ref{tab:1}). The pink contours show results for DESI+Panth.+RSD; the blue contours show results for DESI+Panth.+RSD+$\theta_*$, with $\theta_*$ expressed as "CMB" in the legend.}
\end{figure}

To comment briefly on $\sigma_8$, across Tables \ref{tab:1}, \ref{tab:2}, and \ref{tab:3}, we observe lower values in the $X(z)$CDM analyses than those of $\Lambda$CDM both with and without $\theta_*$. Including $\theta_*$, the $X(z)$CDM value of $\sigma_8 = 0.731 \substack{+0.015\\-0.017}$ in \ref{tab:1} differs from the 2018 reported \textit{Planck} value of $0.811\pm0.006$ at the 4.7$\sigma$ level.

To compare an $X(z)$CDM value of $H_0$ with those from other analyses, we confine the rest of our discussion to the results in Table \ref{tab:1}, reflecting the analysis with the fewest free variables.

First, we consider the results of our analyses excluding $\theta_*$. For $X(z)$CDM, we report $H_0=73.38 \substack{+0.89\\-0.85}$ km/s/Mpc, with $X(z)$ free parameters $x_1 = 0.37 \substack{+0.13\\-0.15} $ and $x_2 = -1.69 \substack{+0.48\\-0.74} $. The corresponding $\Lambda$CDM results are $H_0 = 70.59 \substack{+0.75\\-0.79}$ km/s/Mpc, a deviation of $2.4\sigma$ from the $X(z)$CDM case. 

The addition of $\theta_*$ lowers $H_0$ universally, as expected, giving rise to a larger discrepancy of $3.6\sigma$ between the two models. In the $X(z)$CDM case, $H_0=72.33 \substack{+0.63\\-0.80}$ km/s/Mpc, with $x_1 = 0.42 \substack{+0.14\\-0.13} $ and $x_2 = -1.74\substack{+0.60\\-0.59} $. Even with the addition of condensed CMB information, $H_0$ remains significantly closer to the Cepheid-SNe-Ia deduced value of $73.04 \pm 1.04$ km/s/Mpc \cite{Riess_2022}, differing by only $0.6\sigma$. When compared with the \textit{Planck} value of $H_0=67.4 \pm 0.5$ km/s/Mpc \cite{Planck:2018}, a much larger discrepancy of $5.7\sigma$ arises, confirming that the $X(z)$CDM model fails to resolve the tension between Hubble constant values when constrained by these three datasets and the angular acoustic scale. An interesting extension of this work would be to incorporate additional CMB information beyond $\theta_*$ from \textit{Planck} and the ACT to be used in combination with BAO, SNe-Ia, and RSD data.

\begin{figure}
\centering
\includegraphics[width=0.5\textwidth]{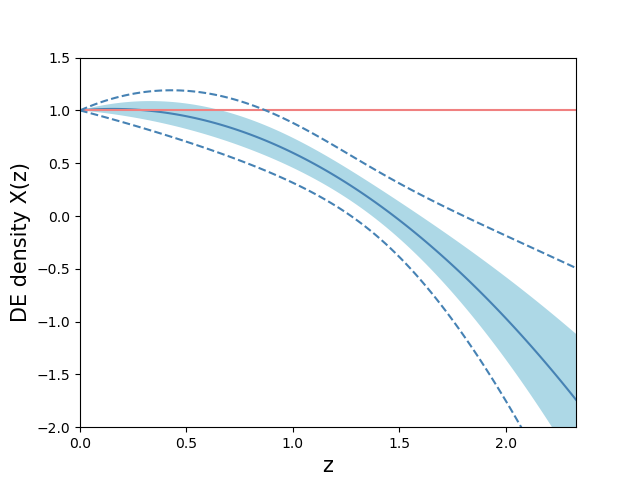}
\caption{\label{fig:X(z)reconstruction} Reconstruction of $X(z)$ using results from Table \ref{tab:1}, where $X(z)$ free parameters are constrained by DESI+Panth.+RSD+$\theta_*$. The blue shaded region represents the $1\sigma$ confidence interval; the blue dashed lines represents the $2\sigma$ confidence interval. The $\Lambda$CDM (cosmological constant) case is shown in pink.}
\end{figure}

The results of the $X(z)$CDM analysis, whilst consistent with model-independent, local $H_0$ determinations, remain in tension with those of \textit{Planck} \cite{Planck:2018}. Hence, we conclude that $X(z)$CDM does not resolve the Hubble Tension, in so far as the model does not rectify the notorious discrepancy between $H_0$ values. Nonetheless, the model may still offer some insight into the nature of DE at low redshift. More specifically, in all $X(z)$CDM cases, the data prefer $0 < x_1 < 1$ and  $ x_2 < -1 $. Both parameters showing significant deviation from unity is suggestive of dynamical DE at low redshift. Using the best fit parameters for $X(z)$ in Table \ref{tab:1}, DE density is re-parametrised and plotted as a function of redshift ($0<z<2.33$), as shown in Fig. \ref{fig:X(z)reconstruction}. As $z$ increases beyond $\sim 0.5$, DE density begins to exhibit dynamical behaviour and decays quadratically, assuming negative values beyond $z\sim 1.5$. DE density decreasing beyond zero is supported at the 2$\sigma$ above the best fit curve. 

In our case, a negative dark energy (DE) density does not violate any physical principle, as we are using this parametrization strictly within a phenomenological framework. What remains entirely prohibited, however, is a negative total density. Actually, models with negative DE density have been considered before. In \cite{Ahmed:2002mj}, the author explored a model inspired by unimodular gravity, which predicts fluctuations in the cosmological constant. These fluctuations are always on the order of the ambient density, making it unsurprising that the cosmological ``constant'' may sometimes take negative values. Additionally, in a similar context, using the back-reaction approach, the author of \cite{Brandenberger:2002sk} found that the cosmological constant oscillates around half the total density on Hubble time scales. An interesting related study was conducted in \cite{Quartin:2008px}, where the authors examined an interacting dark matter/dark energy model using a moderately general interaction term. One of their key conclusions was that, based on their examples, resolving the coincidence problem would require the DE density to have been negative in the past. This feature was first identified using this reconstruction method in \cite{Cardenas:2014jya}, and similar results were examined in \cite{Wang:2018fng}.

At low redshift, $z \lesssim 0.5$, $X(z)=1$, in agreement with $\Lambda$CDM, although this feature of the $X(z)$ reconstruction should be interpreted with caution. A quadratic parametrisation lacks the ability to detect fluctuations at low redshifts-- it is constrained by the properties of a second degree polynomial. The authors of \cite{gialamas2024interpreting} assert that SNe-Ia data at very low redshift, $z<\mathcal{O}(0.1)$, is responsible for the preference of dynamical dark energy over a cosmological constant. Furthermore, by excluding this low redshift data, they show that no deviation from the $\Lambda$CDM model is necessary. 

This prompts interesting questions about the nature of our local Universe, potentially challenging the cosmological principle (CP) for $z<\mathcal{O}(0.1)$. This is not a novel idea; the authors of \cite{Aluri:2022hzs} provide a comprehensive survey of observational data seemingly at odds with predictions of the CP. Perhaps there exists some local gravitational effect due to anisotropic matter distribution that we are yet to account for. Exploring this phenomenon is not within the scope of our analysis but its mentioning serves to contextualise our results at $z<\mathcal{O}(0.1)$. Hence, we emphasise that $X(z)$ tending to unity as $z$ tends to zero without any perturbations could purely be a consequence of a quadratic parametrisation. 

\begin{figure}
\centering
\includegraphics[width=0.5\textwidth]{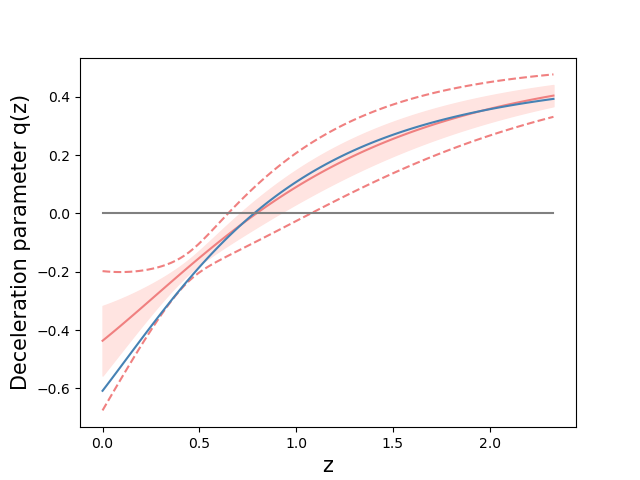}
\caption{\label{fig:deceleration parameter} Reconstruction of $q(z)$ using results from Table \ref{tab:1}, where free parameters are constrained by DESI+Panth.+RSD+$\theta_*$. We show a reconstructed $q(z)$ from $X(z)$CDM results (pink), and the $\Lambda$CDM predicted graph of $q(z)$ (blue). The pink shaded region represents the $1\sigma$ confidence interval; the pink dashed lines represent the $2\sigma$ confidence interval for $X(z)$CDM results.}
\end{figure}

We plot the deceleration parameter, $q(z)$, as a function of redshift in the range $0<z<2.33$, as shown in Fig. \ref{fig:deceleration parameter}. Although the DE density clearly deviates from the $\Lambda$CDM model, even dipping into negative values, the reconstructed deceleration parameter remains consistent with the $\Lambda$CDM model. Notably, the redshift at the deceleration-to-acceleration transition, $z \simeq 0.8$, is identical in both models. Additionally, the current value of the deceleration parameter is slightly lower than the $\Lambda$CDM value, aligning closely with the standard value of $q_0=-0.55$.

\section{Conclusions}

In this paper we explore the implications of using a quadratic DE density parametrisation, $X(z)$, in place of a cosmological constant. With MCMC sampling, we analyse the effectiveness of the $X(z)$CDM model at describing the latest DESI BAO data in combination with SNe-Ia and RSD samples. Using reduced $\chi^2$ and the DW statistic to evaluate goodness of fit, we find no significant preference for $X(z)$CDM over $\Lambda$CDM. We note that both models perform very well according to these statistical measures. Further, we consider the effect on resulting cosmological parameters of constraining the parameter space with the angular acoustic scale. The addition of CMB information in this form lowers $H_0$ in all analyses. 

We present a fundamental way of overcoming the $r_\text{d}$-$H_0$ degeneracy problem in the analysis of BAO data, using an integral equation for $r_\text{d}$. This method, nonetheless, is still dependent on external knowledge from early-time physics predictions and measurements. We consider the consequences of increasing uncertainty in this external information, finding that for $\Lambda$CDM, this has a notable impact on best fit parameters. Specifically, best fit values of $H_0$ increase with increased variability. We do not observe this trend in the $X(z)$CDM results.

Finally, our $X(z)$CDM analysis is indicative dark energy evolution at low redshift: in all cases, $x_1$ and $x_2$ deviate from unity. From our reconstruction of $X(z)$, we find that from $z \sim 0.5$ onwards, DE exhibits dynamical behaviour, and proceeds to assume negative values beyond redshift $\sim 1.5$.

\section*{Acknowledgments}
LO would like to acknowledge Daniel Scolnic for his mentorship. VHC would like to thank CEFITEV-UV for partial support.

\bibliographystyle{apsrev4-1}
\bibliography{references}

\end{document}